\documentclass[runningheads]{llncs}
\usepackage{epsfig}
\usepackage{graphicx}
\usepackage{amsmath}
\usepackage{amssymb}
\usepackage[british]{babel}
\usepackage{csquotes}
\usepackage{graphicx}
\usepackage{subcaption}
\usepackage{booktabs}
\usepackage{physics}
\usepackage{amsmath}

\usepackage{tabu}
\usepackage{multirow, makecell}
\newcommand\T{\rule{0pt}{2.9ex}}       
\newcommand\B{\rule[-1.2ex]{0pt}{0pt}} 

\usepackage[pagebackref=true,breaklinks=true,colorlinks,bookmarks=false]{hyperref}

\usepackage{cleveref}

\begin{document}


\title{Detecting Melanoma Fairly:\\Skin Tone Detection and Debiasing for Skin Lesion Classification}

\titlerunning{Detecting Melanoma Fairly}



\author{Peter J. Bevan\inst{1} \and
Amir Atapour-Abarghouei\inst{2}}

\authorrunning{P. Bevan and A. Atapour-Abarghouei.}

\institute{School of Computing, Newcastle University, UK \email{peterbevan@hotmail.co.uk} \and
Department of Computer Science, Durham University, UK
}

\setlength{\abovedisplayskip}{1mm}
\setlength{\belowdisplayskip}{1mm}

\maketitle

\begin{abstract}

Convolutional Neural Networks have demonstrated human-level performance in the classification of melanoma and other skin lesions, but evident performance disparities between differing skin tones should be addressed before widespread deployment. In this work, we propose an efficient yet effective algorithm for automatically labelling the skin tone of lesion images, and use this to annotate the benchmark ISIC dataset. We subsequently use these automated labels as the target for two leading bias \enquote{unlearning} techniques towards mitigating skin tone bias. Our experimental results provide evidence that our skin tone detection algorithm outperforms existing solutions and that \enquote{unlearning} skin tone may improve generalisation and can reduce the performance disparity between melanoma detection in lighter and darker skin tones.

\end{abstract}

\section{Introduction}
Convolutional Neural Networks (CNN) have demonstrated impressive performance on a variety of medical imaging tasks, one such being the classification of skin lesion images \cite{haenssleManMachineDiagnostic2018a,brinkerConvolutionalNeuralNetwork2019,brinkerComparingArtificialIntelligence2019}.
However, there are also many potential pitfalls that must be identified and mitigated before widespread deployment to prevent the replication of mistakes and systematic issues on a massive scale. For example, an issue that is commonly raised in the existing literature is skin tone bias in lesion classification tasks. Groh et al. \cite{grohEvaluatingDeepNeural2021} provide a compiled dataset of clinical lesions with human annotated Fitzpatrick skin type \cite{fitzpatrickValidityPracticalitySunReactive1988} labels, and show that CNNs perform best at classifying skin types similar to the skin types in the training data used. We use the skin type labels in this dataset as the target for supervised debiasing methods to evaluate the effectiveness of these methods at helping melanoma classification models generalise to unseen skin types.

\begin{figure}
\includegraphics[width=0.5\linewidth]{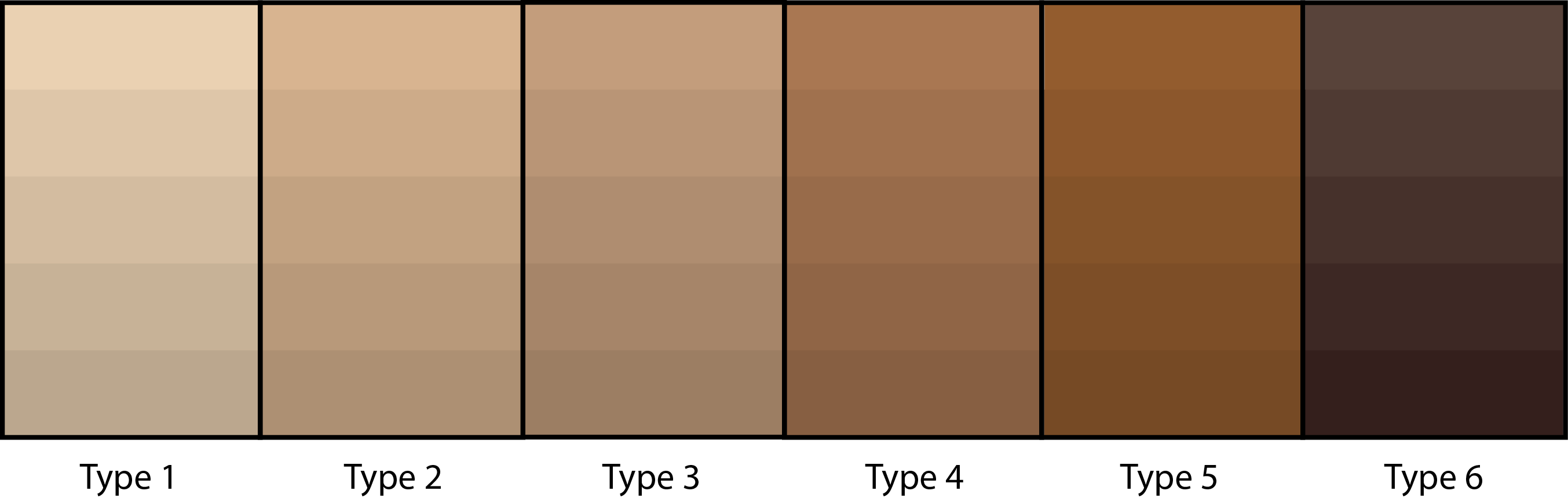}
\centering
\caption{Visualisation of the Fitzpatrick 6 point scale \cite{fitzpatrickValidityPracticalitySunReactive1988}, widely accepted as the gold standard amongst dermatologists \cite{buolamwiniGenderShadesIntersectional2018a}.}
\label{fig:fitz_fig}
\end{figure}

Once we have evaluated the effectiveness of the debiasing methods using human labelled skin tone labels, we look to automate the pipeline further, since human annotated labels are expensive and impractical to gather in practice. We use a novel variation on the skin tone labelling algorithm presented in \cite{kinyanjuiEstimatingSkinTone2019} to annotate the ISIC data and subsequently use these generated labels as the target for a debiasing head, towards creating a fully automated solution to improving the generalisation of models to images of individuals from differing ethnic origins.

In summary, our primary contributions towards the discussed issues are:
\begin{itemize}

  \item \textit{Skin tone detection} - We propose an effective skin tone detection algorithm inspired by \cite{kinyanjuiEstimatingSkinTone2019} (Section \ref{subsec:results:labelling}), the results of which can be used as labels for skin tone bias removal.
  \item \textit{Skin tone debiasing} - We assess the effectiveness of leading debiasing methods \cite{kimLearningNotLearn2019a,alviTurningBlindEye2019} for skin tone bias removal in melanoma classification, and implement these using automated labels as the target for debiasing (Sections \ref{subsec:results:fitz_debias} and \ref{subsec:results:isic_debias}).
\end{itemize}
Code is available at \href{https://github.com/pbevan1/Detecting-Melanoma-Fairly}{https://github.com/pbevan1/Detecting-Melanoma-Fairly}.

\section{Related work}
\label{sec:related}

Groh et al. \cite{grohEvaluatingDeepNeural2021} illustrate that CNNs perform better at classifying images with similar skin tones to those the model was trained on. Performance is, therefore, likely to be poor for patients with darker skin tones when the training data is predominantly images of light-skinned patients, which is the case with many of the current commonly-used dermoscopic training datasets such as the ISIC archive data \cite{rotembergPatientcentricDatasetImages2021,codellaSkinLesionAnalysis2018}. While melanoma incidence is much lower among the black population (1.0 per 100,000 compared to 23.5 per 100,000 for whites), 10-year melanoma-specific survival is lower for black patients (73\%) than white patients (88\%) or other races (85\%) \cite{collinsRacialDifferencesSurvival2011}, and so it is of heightened importance to classify lesions in patients of colour correctly.

One way to ensure a more even classification performance across skin tones is to re-balance the training data by collecting more high-quality images of lesions on skin of colour, but the low incidence of melanoma in darker skin means this could be a slow process over many years. During the time that unbalanced data continues to be an issue, a robust automated method for removing skin tone bias from the model pipeline could potentially help models to operate with increased fairness across skin tones.

\section{Methods}
\label{sec:methods}

\subsection{Debiasing methods}

\begin{figure}[htbp]
\begin{center}
\includegraphics[width=0.8\linewidth]{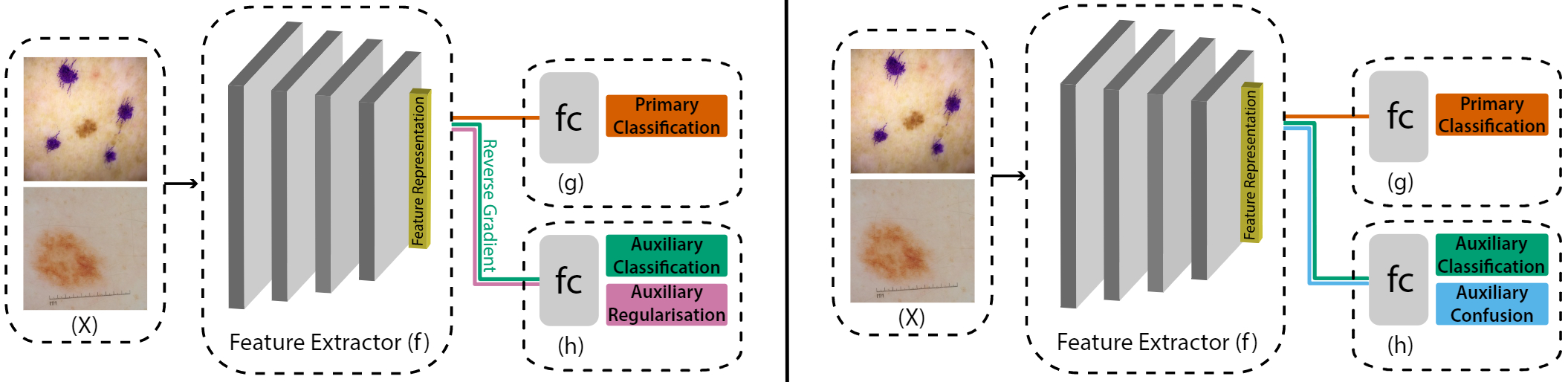}
\centering
\caption{\enquote{Learning Not to Learn} architecture (left) and \enquote{Turning a Blind Eye} architecture (right). $f$ is implemented as a convolutional architecture such as ResNeXt or EfficientNet in this work. \enquote{fc} denotes a fully connected layer.}
\label{fig:LNTL_TABE}
\end{center}
\end{figure}

In this work, two leading debiasing techniques within the literature are used, namely \enquote{Learning Not To Learn} (LNTL) \cite{kimLearningNotLearn2019a} and \enquote{Turning a Blind Eye} (TABE) \cite{alviTurningBlindEye2019}. Both are often referred to as \enquote{unlearning} techniques because of their ability to remove bias from the feature representation of a network by minimising the mutual information between the feature embedding and the unwanted bias. Generic schematics of both \enquote{Learning Not to Learn} and \enquote{Turning a Blind Eye} are shown in \Cref{fig:LNTL_TABE}.

\subsubsection{Learning Not to Learn}
\label{subsec:methods:lntl}
\enquote{Learning Not to Learn} (LNTL) \cite{kimLearningNotLearn2019a} introduces a secondary regularisation loss in combination with a gradient reversal layer \cite{ganinDomainAdversarialTrainingNeural2017} to remove a target bias from the feature representation of a CNN during training. 

The input image, $x$, is passed into a CNN feature extractor, $f$: $x \rightarrow \mathbb{R}^K$, where $K$ is the dimension of the embedded feature. 

The extracted feature embedding is then passed in parallel into the primary classification head $g$: \(\mathbb{R}^K \rightarrow \mathcal{Y}\) and the secondary bias classification head $h$: \(\mathbb{R}^K \rightarrow \mathcal{B}\). \(\mathcal{Y}\) denotes the set of possible lesion classes and \(\mathcal{B}\) denotes the target bias classes.

Formulated as a minimax game, $h$ minimises cross-entropy, learning to classify bias from the extracted features, whilst $f$ maximises cross-entropy, restraining $h$ from predicting the bias, and also minimises negative conditional entropy, reducing the mutual information between the feature representation and the bias. The gradient reversal layer between $h$ and $f$ is used as an additional step to remove information relating to the target bias from the feature representation by multiplying the gradient of the secondary classification loss by a negative scalar during backpropagation, further facilitating the feature extraction network, $f$, to \enquote{unlearn} the targeted bias, $b(x)$. On completion of training, $f$ extracts a feature embedding absent of bias information, $g$ uses this feature embedding to perform an unbiased primary classification, and the performance of $h$ has deteriorated because of the resulting lack of bias signal in the feature embedding.

\subsubsection{Turning a Blind Eye}
\label{subsec:methods:tabe}

\enquote{Turning a Blind Eye} (TABE) also removes unwanted bias using a secondary classifier, $\theta_m$, $m$ being the $m$-th bias to be removed. The TABE secondary classifier identifies bias in the feature representation  $\theta\textsubscript{repr}$ by minimising a secondary classification loss, $\mathcal{L}_s$, and also a secondary confusion loss \cite{tzengSimultaneousDeepTransfer2015a}, $\mathcal{L}\textsubscript{conf}$, which pushes $\theta\textsubscript{repr}$ towards invariance to the identified bias. The losses are minimised in separate steps since they oppose one another: $\mathcal{L}_s$ is minimised alone, followed by the primary classification loss, $\mathcal{L}_p$, together with $\mathcal{L}\textsubscript{conf}$. The confusion loss calculates the cross entropy between a uniform distribution and the output predicted bias. 

As suggested in \cite{kimLearningNotLearn2019a}, TABE can also apply gradient reversal (GR) to the secondary classification loss, and is referred to as \enquote{CLGR} in this work.

\subsection{Skin tone detection}

We calculate the individual typology angle (ITA) of the healthy skin in each image to approximate skin tone \cite{kinyanjuiEstimatingSkinTone2019,grohEvaluatingDeepNeural2021}, given by: 
\begin{equation} \label{eq:ITA}
ITA = arctan\bigg(\frac{L-50}{b}\bigg)\times{\frac{180}{\pi}},
\end{equation}
where $L$ and $b$ are obtained by converting RGB pixel values to the CIELAB colour space. We propose a simpler and more efficient method for isolating healthy skin than the segmentation method used in \cite{kinyanjuiEstimatingSkinTone2019,grohEvaluatingDeepNeural2021}. Across all skin tones, lesions and blemishes are mostly darker than the surrounding skin. Consequently, to select a non-diseased patch of skin, we take 8 samples of 20$\times$20 pixels from around the edges of each image and use the sample with the highest ITA value (lightest skin tone) as the estimated skin tone. The idea behind replacing segmentation with this method is to reduce the impact of variable lighting conditions on the skin tone estimation by selecting the lightest sample rather than the entire healthy skin area. This method is also quicker and more efficient than segmentation methods due to its simplicity.

Eq. \ref{eq:ITA_thresh} shows the thresholds set out in \cite{grohEvaluatingDeepNeural2021}, which are taken from \cite{kinyanjuiEstimatingSkinTone2019} and modified to fit the Fitzpatrick 6 point scale \cite{fitzpatrickValidityPracticalitySunReactive1988} (see \Cref{fig:fitz_fig}). We use these thresholds in our skin tone labelling algorithm.
\begin{equation} \label{eq:ITA_thresh}
\resizebox{0.4\hsize}{!}{$
Fitzpatrick(ITA)=
\begin{cases}
    1   & ITA > 55 \\
    2   & 55 \geq ITA > 41 \\
    3   & 41 \geq ITA > 28 \\
    4   & 28 \geq ITA > 19 \\
    5   & 19 \geq ITA > 10 \\
    6   & 10 \geq ITA \\
\end{cases}
$}
\end{equation}
We pre-process each image using black-hat morphology to remove hair, preventing dark pixels from hairs skewing the calculation. This hair removal is purely for skin tone detection and the original images are used for training the debiased classification models. It is clear that even with large lesions with hard-to-define borders, our method is highly likely to select a sample of healthy skin.
\begin{figure*}
\includegraphics[width=\linewidth]{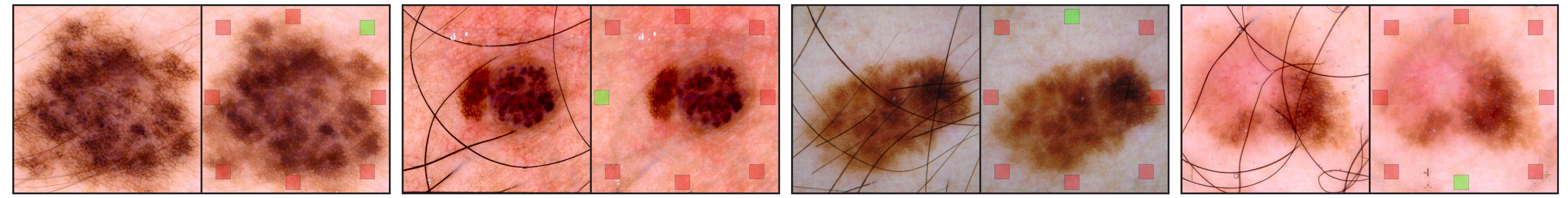}
\centering
\caption{Left of each pair shows ISIC input images, right of each pair shows the placement of the 20$\times$20 pixel samples on images with hair removed. Green square indicates chosen sample based on lightest calculated tone. This sampling method eliminates the need for segmentation.}
\label{fig:detect}

\end{figure*}

\subsection{Data}

\subsubsection{Training data}
A compilation of clinical skin condition images with human annotated Fitzpatrick skin types \cite{fitzpatrickValidityPracticalitySunReactive1988}, called the \enquote{Fitzpatrick17k} dataset \cite{grohEvaluatingDeepNeural2021}, is used for training to demonstrate the effectiveness of unlearning for skin tone debiasing, and to evaluate our automated skin tone labelling algorithm. Of the 16,577 images, we focus on the 4,316 of these that are neoplastic (tumorous). These labels are provided by non-dermatologist annotators, so are likely to be imperfect. When attempting dibiasing of ISIC data, a combination of the 2017 and 2020 challenge data \cite{rotembergPatientcentricDatasetImages2021,codellaSkinLesionAnalysis2018} (35,574 images) is used as training data.

\subsubsection{Test data}
The MClass \cite{brinkerComparingArtificialIntelligence2019} dataset is used to evaluate generalisation and provide a human benchmark. This dataset comprises a set of 100 dermoscopic images and 100 clinical images (\textit{different} lesions), each with 20 malignant and 80 benign lesions. The human benchmark is the classification performance of 157 dermatologists on the images in the dataset. The Interactive Atlas of Dermoscopy \cite{lioInteractiveAtlasDermoscopy2004}, and the ASAN datasets \cite{hanClassificationClinicalImages2018} were used to further test the robustness of the models. The Atlas dataset has 1,000 lesions, with one dermoscopic and one clinical image per lesion (2,000 total), while the ASAN test dataset has 852 images, all clinical. Whilst the ISIC training data \cite{rotembergPatientcentricDatasetImages2021,codellaSkinLesionAnalysis2018} is mostly white Western patients, the Atlas seems to have representation from a broad variety of ethnic groups, and ASAN from predominantly South Korean patients, which should allow for a good test of a model's ability to deal with different domain shifts.

\subsection{Implementation}

PyTorch \cite{NEURIPS2019_9015} is used to implement the models. The setup used for experimentation consists of two NVIDIA Titan RTX GPUs in parallel with a combined memory of 48 GB on an Arch Linux system with a 3.30GHz 10-core Intel CPU and 64 GB of memory. The source code is publicly released to enable reproducibility and further technical analysis.

After experimentation with EfficientNet-B3 \cite{tanEfficientNetRethinkingModel2019}, ResNet-101 \cite{heDeepResidualLearning2016}, ResNeXt-101 \cite{xieAggregatedResidualTransformations2017}, DenseNet \cite{huangDenselyConnectedConvolutional2017} and Inception-v3 \cite{szegedyRethinkingInceptionArchitecture2016a}, ResNeXt-101 looked to show the best performance and so was used as the feature extractor in the debiasing experiments. All classification heads are implemented as one fully-connected layer, as in \cite{kimLearningNotLearn2019a}.
Stochastic gradient descent (SGD) is used across all models, ensuring comparability and compatibility between the baseline and debiasing networks.

\section{Experimental results}
\label{sec:results}

\subsection{Fitzpatrick17k skin tone debiasing}
\label{subsec:results:fitz_debias}

\begin{table}[b]
	\centering
	\resizebox{0.37\linewidth}{!}{
		{\tabulinesep=0mm
			\begin{tabu}{@{\extracolsep{4pt}}c c c@{}}
				\hline\hline
				Experiment & Types 3\&4 & Types 5\&6\T\B \\
				\cline{2-3}
				\hline\hline
 Baseline & 0.872 & 0.835\T\\
 
 CLGR & \textbf{0.883} & \textbf{0.853}\B\\ 
\hline
\hline
\end{tabu}
}
}
\captionsetup[table]{skip=7pt}
\captionof{table}{Improving model generalisation to skin tones different to the training data \cite{grohEvaluatingDeepNeural2021}. All scores are \textbf{AUC}. Trained using types 1\&2 skin images from the Fitzpatrick17k dataset \cite{grohEvaluatingDeepNeural2021}, tested on types 3\&4 skin and types 5\&6.}
\label{tab:FitzTable}
\end{table}

A CNN trained using Fitzpatrick \cite{fitzpatrickValidityPracticalitySunReactive1988} types 1 and 2 skin is shown to perform better at classifying skin conditions in types 3 and 4 than types 5 and 6 skin in \cite{grohEvaluatingDeepNeural2021}. We are able to reproduce these findings with our baseline ResNeXt-101 model, trained and tested on the neoplastic subset of the Fitzpatrick17k data. Our objective is to close this gap with the addition of a secondary debiasing head which uses skin type labels as its target. The CLGR configuration proves to be most effective, and is shown in \Cref{tab:FitzTable}. The disparity in AUC between the two groups is closed from 0.037 to 0.030, with types 3 and 4 boosted by 1.3\% and types 5 and 6 boosted by 2.2\%. It is important to note that due to the critical nature of the problem and the significant ramifications of false predictions in real-world applications, even small improvements are highly valuable. This experiment serves as a proof of concept for the mitigation of skin tone bias with unlearning techniques, and gives us precedent to explore this for debiasing the ISIC \cite{rotembergPatientcentricDatasetImages2021,codellaSkinLesionAnalysis2018} or other similar datasets. Since the ISIC data does not have human annotated skin tone labels, to explore debiasing this dataset we first generate these labels with an automated skin tone labelling algorithm (see \cref{subsec:results:labelling}).

\subsection{Automated skin tone labelling algorithm}
\label{subsec:results:labelling}

To validate the effectiveness of our skin tone labelling algorithm, we re-label the Fitzpatrick17k data and compare these automated labels against the human annotated skin tones to calculate accuracy, with a correct prediction being within $\pm$1 point on the Fitzpatrick scale \cite{grohEvaluatingDeepNeural2021}. Our method achieves 60.61\% accuracy, in comparison to the 53.30\% accuracy achieved by the algorithm presented in \cite{grohEvaluatingDeepNeural2021}, which segments the healthy skin using a YCbCr masking algorithm. 
The authors of \cite{grohEvaluatingDeepNeural2021} improve their accuracy to 70.38\% using empirically selected ITA thresholds, but we decide against using these to label the ISIC data, given that they are optimised to suit only the Fitzpatrick17k data and do not generalise. 

\begin{figure}[b]
\includegraphics[width=0.5\linewidth]{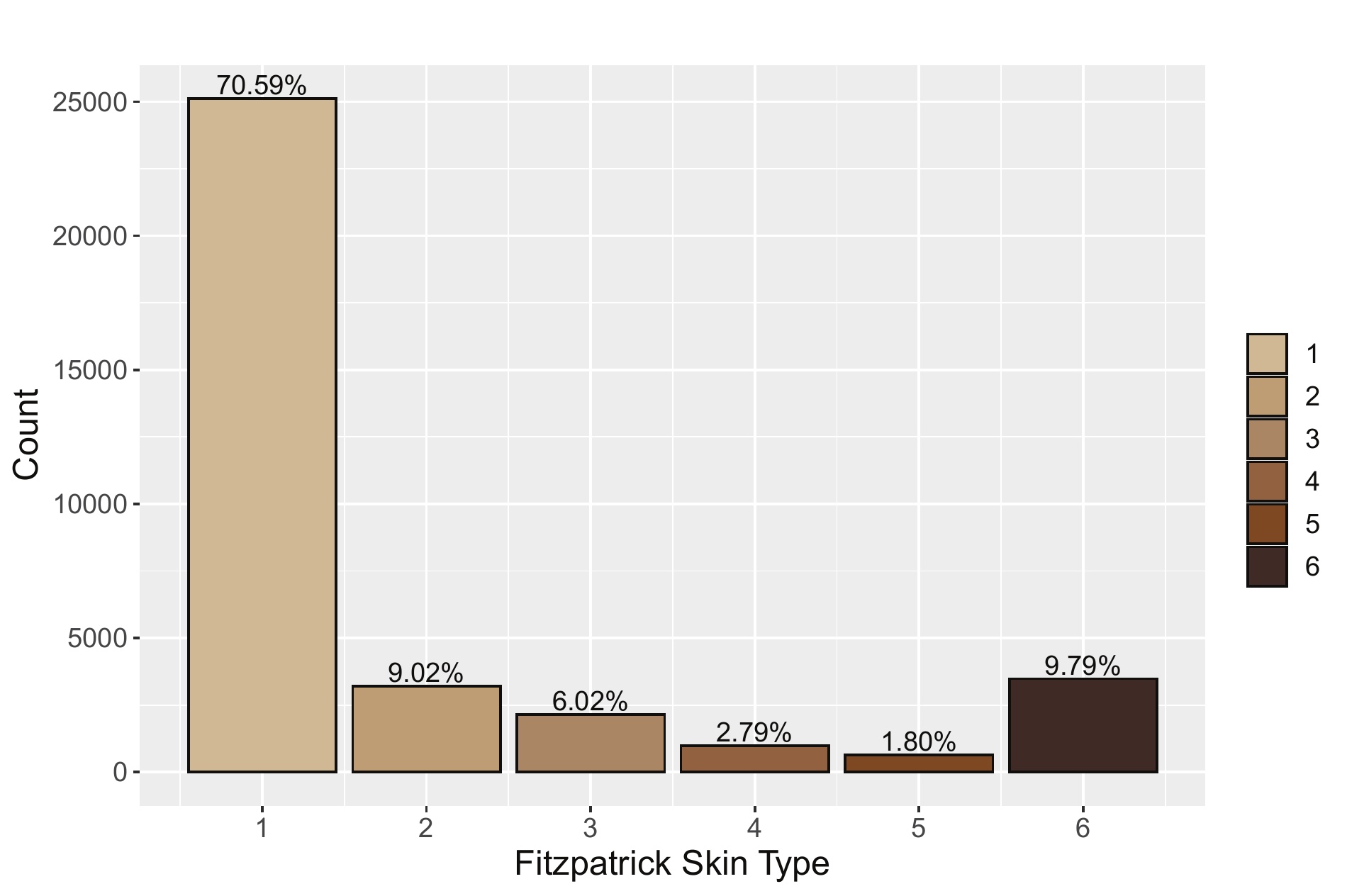}
\centering
\caption{Distribution of Fitzpatrick skin types in ISIC \cite{rotembergPatientcentricDatasetImages2021,codellaSkinLesionAnalysis2018} training data, as labelled by our algorithm.}
\label{fig:isic_dist_fitz}
\end{figure}

We expect our algorithm to perform better still on the ISIC data \cite{rotembergPatientcentricDatasetImages2021,codellaSkinLesionAnalysis2018} than the Fitzpatrick17k data \cite{grohEvaluatingDeepNeural2021}, since the images are less noisy, meaning the assumption that the lightest patch in the image is healthy skin is less likely to be undermined by artefacts or a lightly coloured background.

\Cref{fig:isic_dist_fitz} shows the distribution of Fitzpatrick skin types in the ISIC training data, labelled by our skin tone detection algorithm. The figure shows a clear imbalance towards lighter skin tones. The relatively high number of type 6 classifications could be due to the labelling algorithm picking up on dark lighting conditions, since upon visual inspection of the dataset, it can be concluded that there is not likely to be this many type 6 skin images in the dataset. This is something that should be explored and improved in future work.

\subsection{ISIC skin tone debiasing}
\label{subsec:results:isic_debias}

The ISIC archive is one of the most popular publicly available melanoma training datasets, but there are no skin tone labels available, so we use our skin tone labelling algorithm to analyse the distribution of skin tones in this data as well as to further test the debiasing methods. We also use these labels as the target for the debiasing heads during training. Although these labels have low accuracy, it has been shown that deep learning is still able to learn, even in cases where labels are noisy \cite{jiang_beyond_2020}. We see a small performance improvement across the board when debiasing with the TABE \cite{alviTurningBlindEye2019} head, indicating that this model generalises to the test sets better than the baseline (see \Cref{tab:ISICRacial}), including a 5.3\% improvement in AUC on the ASAN test set. Performance on this dataset is of particular interest since these images are known to be from Korean patients and so represent a definitive domain shift in comparison to the predominantly Western ISIC training data. The TABE head also prompts a 14.8\% increase in performance on the Atlas clinical test set \cite{lioInteractiveAtlasDermoscopy2004} compared to the baseline, and all debiasing heads show noticeable improvements on the MClass dermoscopic and clinical test sets \cite{brinkerComparingArtificialIntelligence2019}. Although the origins of the Atlas and MClass clinical data are unknown, these also look to be drawn from significantly different populations to the ISIC data (containing many more examples of darker skin tones), so improvements on these test sets could be interpreted as evidence of the mitigation of skin tone bias.

Our models demonstrate superior classification performance compared to the group of dermatologists from \cite{brinkerComparingArtificialIntelligence2019}. While impressive, this comparison should be taken with a grain of salt, as these dermatologists were classifying solely using images and no other information. A standard clinical encounter with each patient would likely result in better performance than this. Moreover, systems like this are not meant to replace the expertise of a dermatologist at this stage, but to augment and enhance the diagnosis and facilitate easier access to certain patients.

\begin{table*}[htbp]
	\centering
	\resizebox{0.77\linewidth}{!}{
		{\tabulinesep=0mm
			\begin{tabu}{@{\extracolsep{12pt}}c c c c c c@{}}
				\hline\hline
				\multicolumn{1}{c}{\multirow{2}{*}{Experiment}} & 
				\multicolumn{2}{c}{Atlas} &
				\multicolumn{1}{c}{Asan} &
				\multicolumn{2}{c}{MClass}\T\B \\
				\cline{2-3} \cline{4-4} \cline{5-6}
						& Dermoscopic & Clinical & Clinical & Dermoscopic & Clinical\T\B\\
				\hline\hline
 Dermatologists & --- & --- & --- & 0.671 & 0.769\T\\

 Baseline & 0.819 & 0.616 & 0.768 & 0.853 & 0.744 \\ 
 
 LNTL & 0.803 & 0.608 & 0.765 & 0.858 & 0.787 \\

 TABE & \textbf{0.825} & \textbf{0.707} & \textbf{0.809} & 0.865 & \textbf{0.859} \\

 CLGR & 0.820 & 0.641 & 0.740 & \textbf{0.918} & 0.771\B\\
\hline
\hline
\end{tabu}
}
}
\captionsetup[table]{skip=7pt}
\captionof{table}{Comparison of skin tone debiasing techniques, with AUC used as the primary metric. Models are trained using ISIC 2020 \& ISIC 2017 data \cite{rotembergPatientcentricDatasetImages2021,codellaSkinLesionAnalysis2018}.}
\label{tab:ISICRacial}
\end{table*}

\subsection{Ablation studies}
\label{subsec:ablation}

TABE \cite{alviTurningBlindEye2019} with and without gradient reversal has provided impressive results, but ablation of the gradient reversal layer from LNTL \cite{kimLearningNotLearn2019a} led to degraded performance (see \Cref{tab:ablate_grl}). Deeper secondary heads were experimented with (additional fully-connected layer), but did not have a noticeable impact on performance (see supplementary material).

\begin{table}
	\centering
	\resizebox{0.37\linewidth}{!}{
		{\tabulinesep=0mm
			\begin{tabu}{@{\extracolsep{4pt}}c c c@{}}
				\hline\hline
				Experiment & Types 3\&4 & Types 5\&6\T\B \\
				\cline{2-3}
				\hline\hline
 LNTL & \textbf{0.873} & \textbf{0.834}\T\\
 
 LNTL* & 0.867 & 0.829\B\\ 
\hline
\hline
\end{tabu}
}
}
\captionsetup[table]{skip=7pt}
\captionof{table}{Ablation of gradient reversal layer from LNTL (ResNext101). Asterisk (*) indicates ablation of gradient reversal).}
\label{tab:ablate_grl}
\end{table}

\section{Limitations and future work}
\label{sec:limitations}

As mentioned in \cref{subsec:results:labelling}, the skin tone detection algorithm has a problem with over-classifying type 6 skin which is a key limitation and should be addressed. ITA is an imperfect method for estimating skin tone, given its sensitivity to lighting conditions, and the Fitzpatrick conversion thresholds are tight and may not generalise well. Empirical calibration of these thresholds tailored to the specific data in question may help, as is done in \cite{grohEvaluatingDeepNeural2021}.

Further work may collect dermatologist annotated skin tone labels for dermoscopic datasets and evaluate the effectiveness of debiasing techniques using these human labels. These labels would also allow a more robust evaluation of skin tone bias in the ISIC data than we were able to provide.

Although this work provides potential methods for bias mitigation in melanoma detection, we caution against over-reliance on this or similar systems as silver bullet solutions, as this could further lead to the root cause of the problem (imbalance and bias within the data) being overlooked. We encourage a multifaceted approach to solving the problem going forward. Further work may also look to do a deeper analysis into the debiasing methods to confirm that the improved generalisation is a result of mitigation of the targeted bias.

\section{Conclusion}
\label{sec:conclusions}

This work has provided evidence that the skin tone bias shown in \cite{grohEvaluatingDeepNeural2021} can be at least partially mitigated by using skin tone as the target for a secondary debiasing head. We have also presented an effective variation of Kinyanjui et al.'s skin tone detection algorithm \cite{kinyanjuiEstimatingSkinTone2019}, and used this to label ISIC data. We have used these labels to unlearn skin tone when training on ISIC data and demonstrated some improvements in generalisation, especially when using a \enquote{Turning a Blind Eye} \cite{alviTurningBlindEye2019} debiasing head. Given that current publicly available data in this field is mostly collected in Western countries, generalisation and bias removal tools such as these may be important in ensuring these models can be deployed to less represented locations as soon as possible in a fair and safe manner.

\small
\bibliographystyle{plain}
\bibliography{Detecting_Melanom_Fairly_MICCAI.bib}
\clearpage

\appendix
\section*{Supplementary Material}
\normalsize

Table 4 shows full results on the Fitzpatrick17k dataset from Section 4.1 of the main paper. We try using auxiliary classification heads with an additional fully connected layer to see if this improves performance, but there is no conclusive answer to from the resulting data and so we stick with the simpler option of using a single fully connected layer.

\begin{figure*}[htbp]
  \begin{subfigure}[b]{0.48\linewidth}
\centering
    \includegraphics[width=\linewidth]{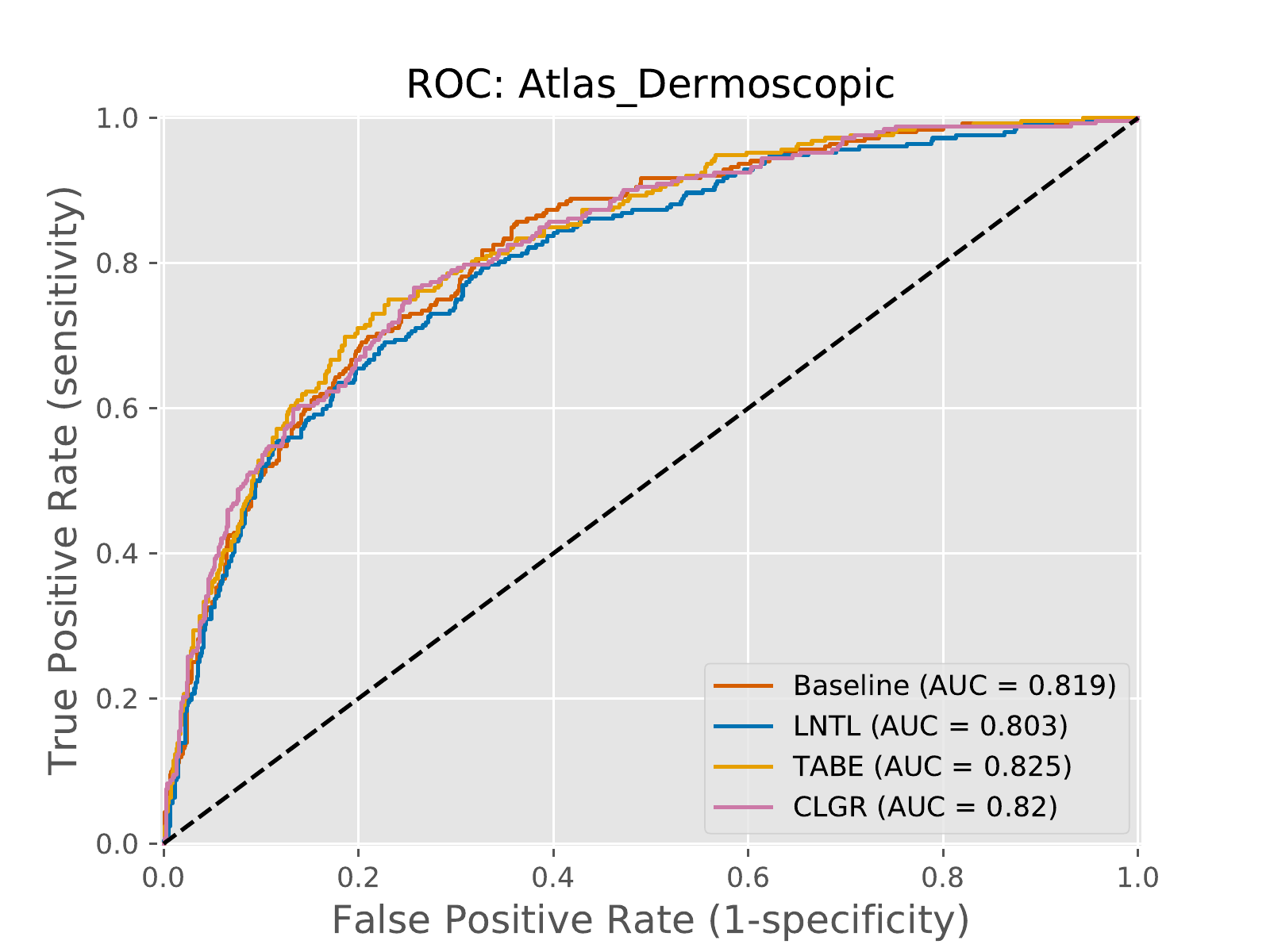}
    \caption{Atlas dermscopic}
  \end{subfigure}
  \begin{subfigure}[b]{0.48\linewidth}
\centering
    \includegraphics[width=\linewidth]{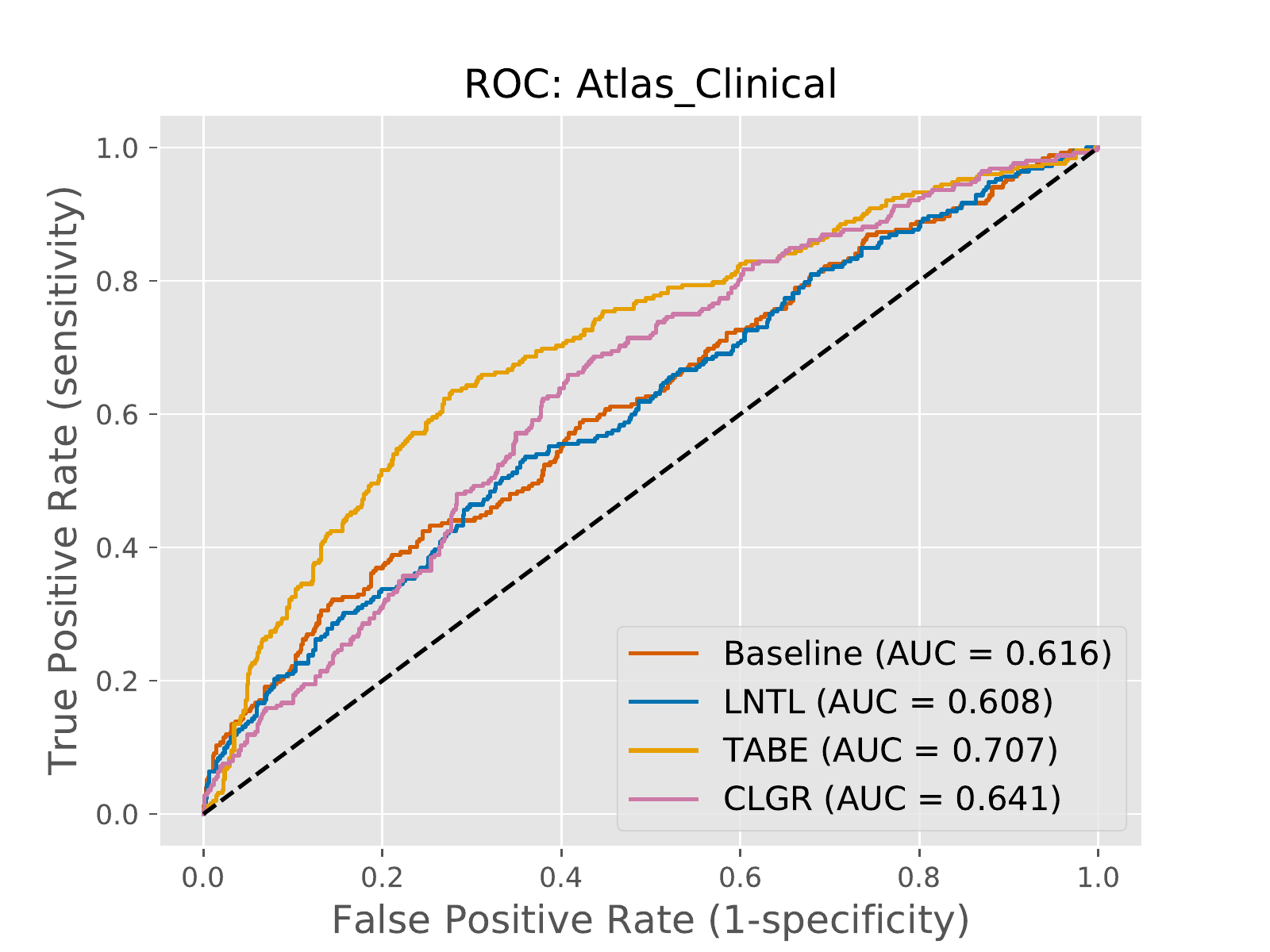}
    \caption{Atlas clinical}
  \end{subfigure}
  \begin{subfigure}[b]{0.48\linewidth}
\centering
    \includegraphics[width=\linewidth]{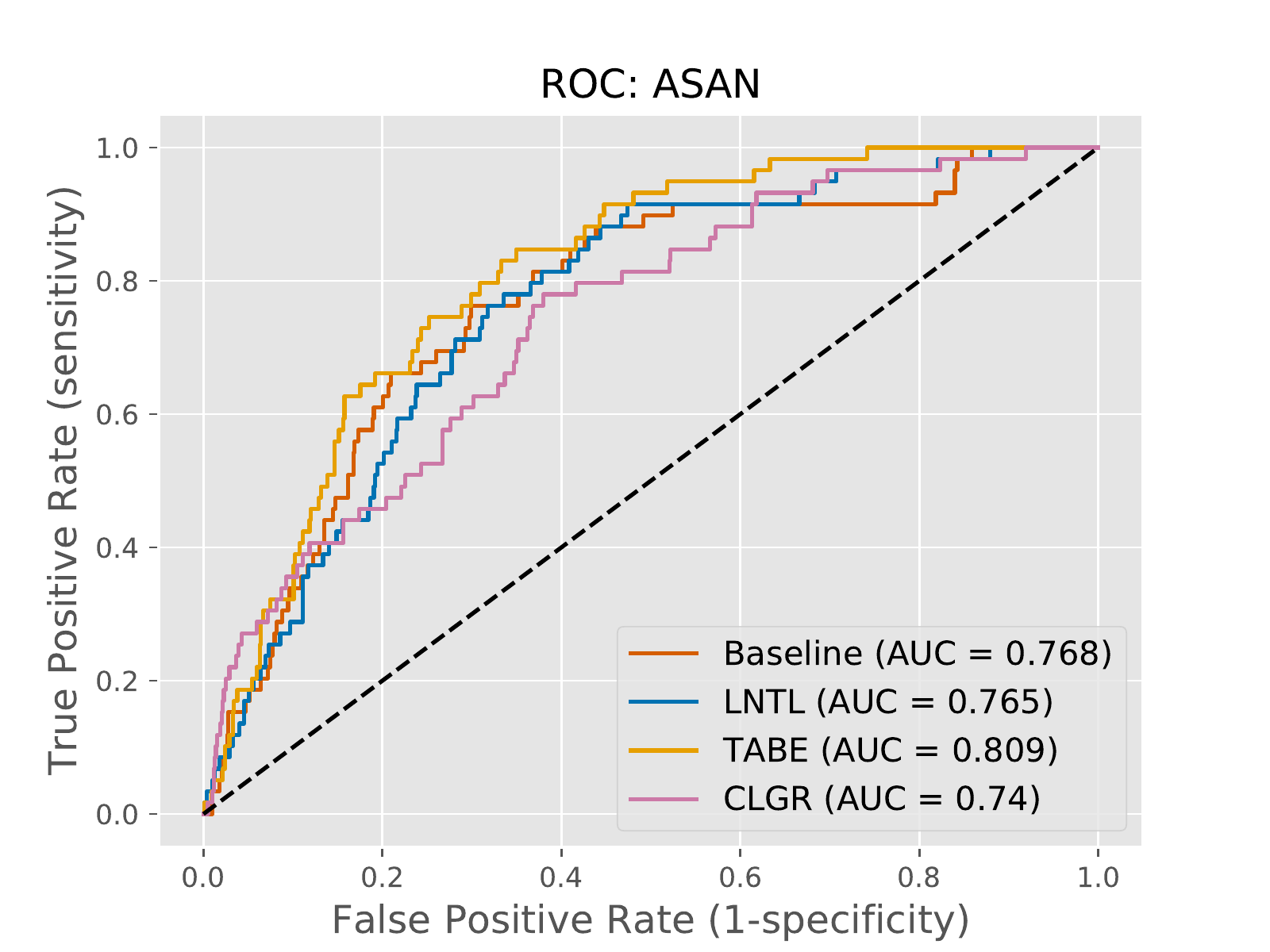}
    \caption{ASAN clinical}
  \end{subfigure}
    \begin{subfigure}[b]{0.48\linewidth}
\centering
    \includegraphics[width=\linewidth]{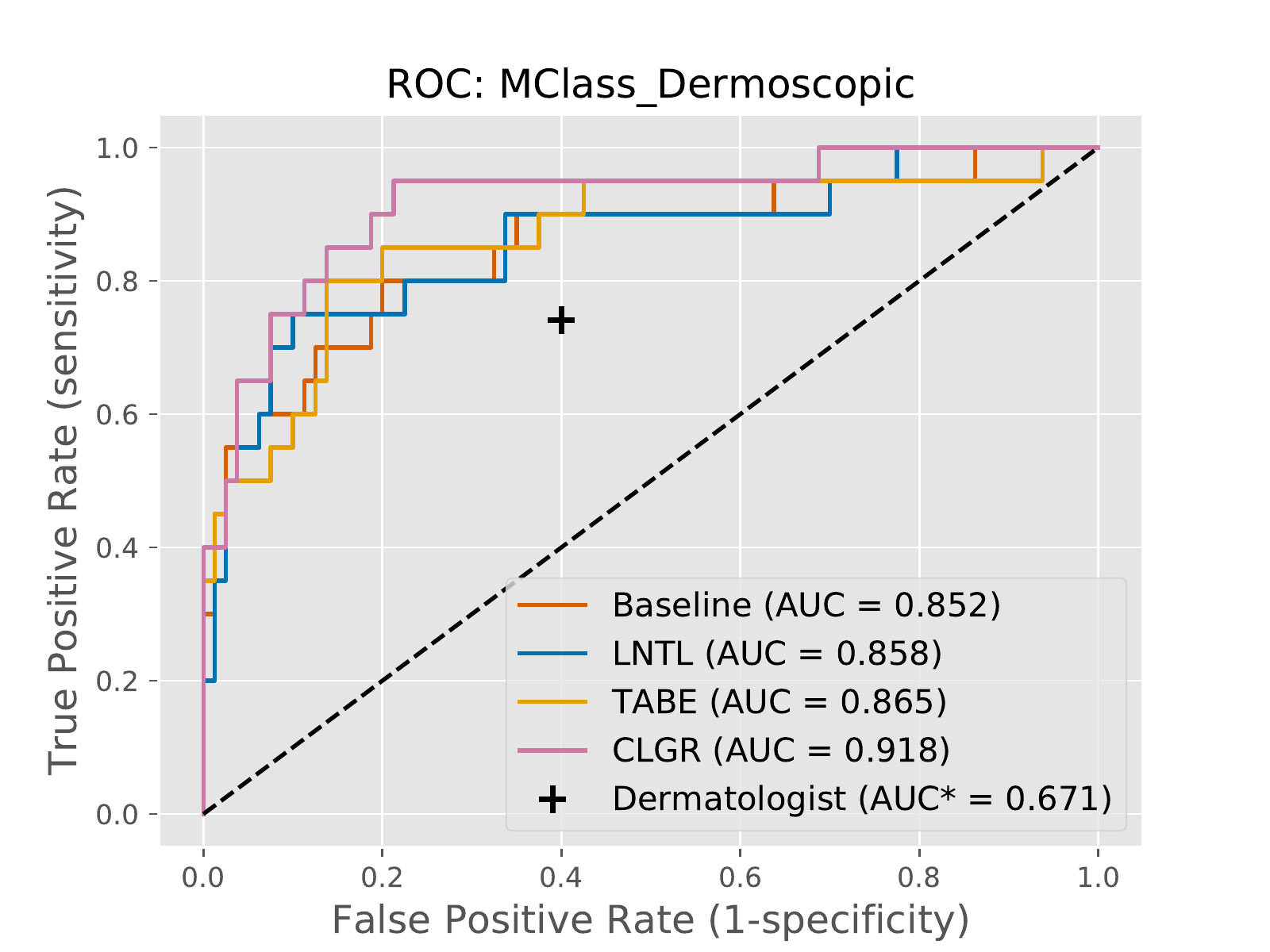}
    \caption{MClass dermoscopic}
  \end{subfigure}
  \begin{subfigure}[b]{0.48\linewidth}
\centering
    \includegraphics[width=\linewidth]{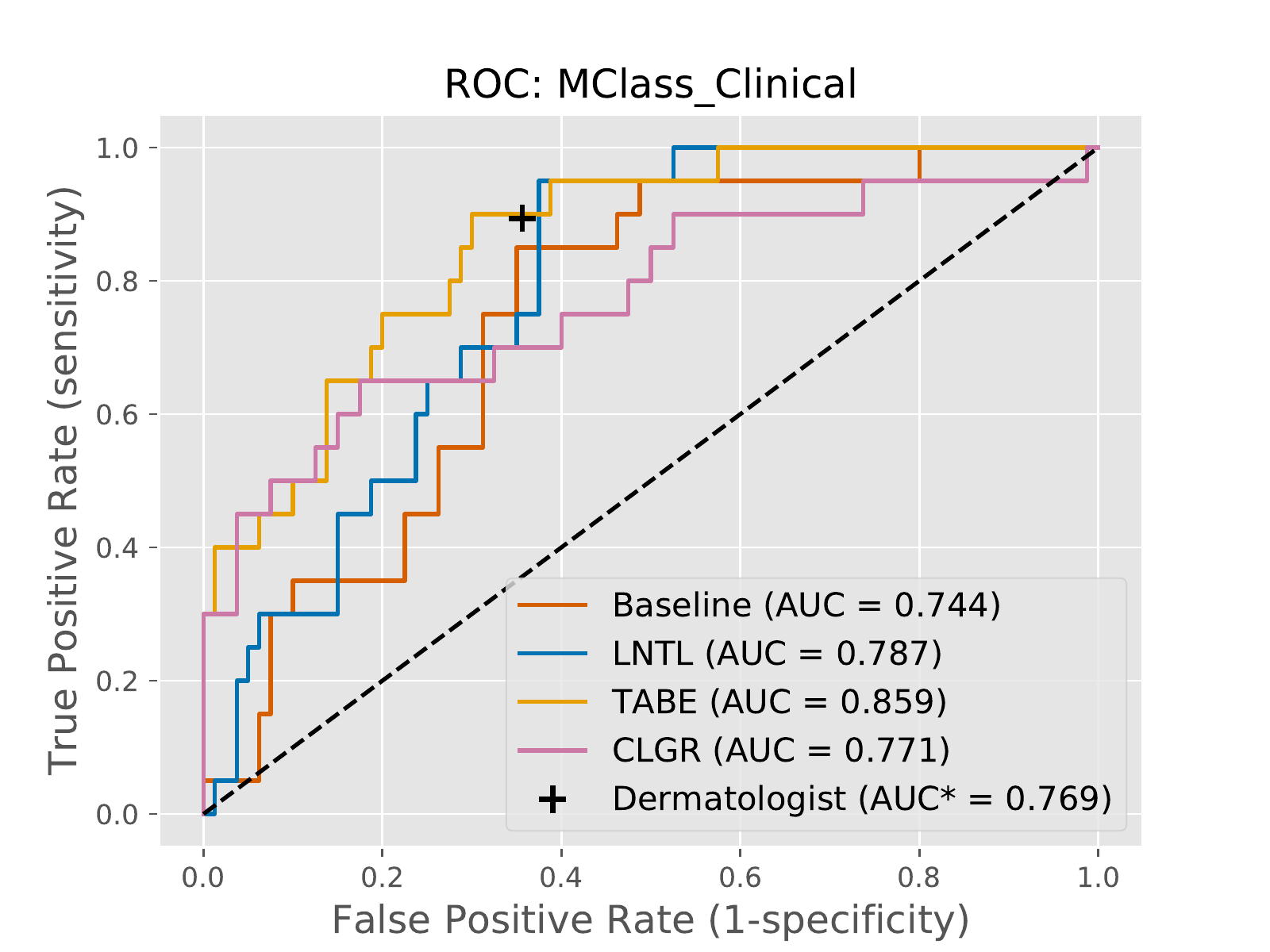}
    \caption{MClass clinical}
  \end{subfigure}
  \caption{ROC curves for each debiasing method, with \textbf{ResNeXt-101} as the base architecture, aiming to remove skin tone bias. Model trained using the ISIC 2020 \cite{rotembergPatientcentricDatasetImages2021} and 2017 data \cite{codellaSkinLesionAnalysis2018}.}
\centering
\label{fig:ROC_sktype_full}
\end{figure*}

\begin{table}[htbp]
	\centering
	\resizebox{0.37\linewidth}{!}{
		{\tabulinesep=0mm
			\begin{tabu}{@{\extracolsep{4pt}}c c c@{}}
				\hline\hline
				Experiment & Types 3\&4 & Types 5\&6\T\B \\
				\cline{2-3}
				\hline\hline
 Baseline & 0.872 & 0.835\T\\

 LNTL & \textbf{0.873} & 0.834 \\ 

 LNTL* & 0.872 & \textbf{0.844} \\ 

 TABE & 0.878 & \textbf{0.849} \\ 

 TABE* & \textbf{0.884} & 0.843 \\ 

 CLGR & 0.883 & \textbf{0.853} \\ 

 CLGR* & \textbf{0.888} & 0.849\B\\
\hline
\hline
\end{tabu}
}
}
\captionsetup[table]{skip=7pt}
\captionof{table}{Attempting to improve model generalisation to skin tones different to the training data \cite{grohEvaluatingDeepNeural2021}. Trained using types 1 and 2 skin images from the Fitzpatrick17k dataset \cite{grohEvaluatingDeepNeural2021}, tested on types 3\&4 and 5\&6 from the same set. Asterisk (*) indicates use of \textbf{deeper head} (additional fully connected layer).}
\label{tab:FitzTable2}
\end{table}

\clearpage

\end{document}